\documentclass[a4paper]{article}
\pdfoutput=1
\usepackage{INTERSPEECH2020}
\usepackage{listings}
\usepackage{url}
\usepackage{comment}
\usepackage{float}

\newcommand{\code}[1]{\texttt{#1}}

\title{Diarization of Legal Proceedings \\ \large Identifying and Transcribing Judicial Speech from Recorded Court Audio}
\name{Jeffrey Tumminia, Amanda Kuznecov, Sophia Tsilerides, Ilana Weinstein, Brian McFee, Michael Picheny, Aaron R. Kaufman}
\address{New York University}
\email{\{jt2565, anr431, smt570, igw212, bm106, map22, ak8096\}@nyu.edu }

\begin{document}
\maketitle
\begin{abstract}
United States Courts make audio recordings of oral arguments available as public record, but these recordings rarely include speaker annotations. This paper addresses the {\it Speech Audio Diarization} problem, answering the question of \textit{"Who spoke when?"} in the domain of judicial oral argument proceedings. We present a workflow for diarizing the speech of judges using audio recordings of oral arguments, a process we call \textit{Reference-Dependent} Speaker Verification. We utilize a speech embedding network trained with the Generalized End-to-End Loss to encode speech into d-vectors and a pre-defined reference audio library based on annotated data. We find that by encoding reference audio for speakers and full arguments and computing similarity scores we achieve a $13.8\%$ Diarization Error Rate for speakers covered by the reference audio library on a held-out test set. We evaluate our method on the Supreme Court of the United States oral arguments, accessed through the \textit{Oyez Project}, and outline future work for diarizing legal proceedings. A code repository for this research is available at \url{github.com/JeffT13/rd-diarization}. 
\end{abstract}

\section{Introduction}

Courts of record across the United States have increasingly migrated to audio recordings as the primary format for the public dissemination of oral arguments. While audio recordings enable widespread access to important records, their utility is limited without transcription and diarization. This is true for both the academic research community and the legal industry: easy access to clean court proceeding text would both facilitate new academic research into the judiciary and contribute to new tools to improve access to affordable legal services.

Diarization answers the question of \textit{"Who spoke when?"}. Common approaches to the problem are derived from the speaker verification literature, and can be divided into either \textit{Text-Dependent} Speaker Verification (TDSV) or \textit{Text-Independent} Speaker Verification (TISV) \cite{wan2020generalized}. Diarization of untranscribed legal proceedings is a TISV task, which is often more difficult than TDSV. Diarization is an open problem in speech processing; casual inspection of a variety of available commercial APIs for this purpose do not produce satisfactory results on legal audio, which is characterized by many unique speakers in a single recording, unequal durations of speech turns both within and across speakers, and unusually long audio recordings. But while these features render court proceeding diarization challenging for existing systems, there are important characteristics of these data that can facilitate diarization as well. In particular, the audio contains minimal background noise and overlapped speech. In addition, due to the judicial system's rich metadata, we know \textit{ex ante} which speakers can be present in a given recording. Since most of the speakers for a given court system reoccur frequently, this renders our task as an open-set Speaker Identification problem \cite{sr_hansen}. We therefore look to implement a diarization pipeline that overcomes the challenging aspects of legal proceedings and exploits the advantageous ones. 

To prototype and evaluate diarization strategies for the legal domain, we leverage data from the \textit{Oyez Project} \cite{oyez}. Oyez conveniently provides a publicly available database of thousands of transcribed, diarized, and time-stamped Supreme Court of the United States (SCOTUS) oral arguments retrievable through a public API. Each proceeding contains at least two lawyers in addition to some or all of the nine Supreme Court Justices. 

This dataset is a useful benchmark for diarization in the legal domain. Records are typically between 40 and 80 minutes long, contain 8-13 speakers who speak in varied amounts, and contain mostly adult male voices, making TISV challenging\footnote{Audio that contain a mix of genders is generally easier to diarize.}. These features are relatively unique among the datasets commonly used in diarization evaluation. Typically, diarization datasets emphasize challenges such as  multiple settings (DIHARD \cite{ryant2021dihard}, SITW \cite{sitwMcLaren}), multilingual conversation (CALLHOME\footnote{\url{https://catalog.ldc.upenn.edu/LDC2001S97}}), or background noise (VOICES \cite{richey2018voices}). They do not contain a high percentage of reoccurring speakers, and so do not well represent our use case. The Broadcast News datasets (\cite{sdbn}, \cite{Pallett19981998BN}, \cite{Pallett19971997BN}) have reoccurring speakers, but are small and in general contain very little interjection.

This paper describes the creation of a simple and efficient diarization open-source tool\footnote{\url{github.com/JeffT13/rd-diarization}} that considers the advantages and challenges in the legal domain, which we refer to as \textit{Reference-Dependent} Speaker Verification (RDSV). We utilize a speech embedding network trained with the Generalized End-to-End Loss \cite{wan2020generalized} to encode speech into d-vectors embeddings \cite{Varianidvec}. Next we use the Oyez annotated audio to generate a dictionary of speakers and their associated d-vectors, the \textit{Reference Audio Library} (RAL). Finally, we perform cosine distance scoring \cite{ivec} of the d-vectors from reference audio and target audio to perform diarization. Our embedder draws on the Resemblyzer\footnote{\url{github.com/resemble-ai/Resemblyzer}}  \textit{Voice Encoder}. Our RDSV pipeline achieves a Diarization Error Rate (DER) of $13.8\%$ on a held-out test set for speakers covered by the reference audio library. 

 Section 2 reviews various strategies for TISV. Section 3 details the Voice Encoder, and Section 4 outlines the RDSV process which includes the RAL structure and Speaker Verification calculations. In Section 5 we investigate the separability of the judicial speech in court audio with the Voice Encoder and apply our pipeline to the Oyez SCOTUS data. In Section 6 we discuss the performance of our method and motivate future research on the topic of diarizing court proceedings.

\section{Related Work}

 Text-Independent-Speaker-Verification (TISV) is commonly approached as two distinct tasks: (1) generating informative speaker representations, and (2) diarization strategies which use these representations to infer speaker labels. For a more complete survey of the study of diarization, particularly via deep learning, see Park et al. \cite{park2021review} and Bai et al. \cite{bai2020speaker}.

Speaker representations are typically extracted embeddings from one of the end layers of a NN trained to perform speaker verification or identification. These typically come in one of two variations, d-vectors (based on DNNs) \cite{Varianidvec} and x-vectors \cite{xvec}. x-vectors extract embeddings from a time-delayed neural network with an intermediate pooling layer, trained to discriminate speakers from variable length segments. Relative to d-vectors, they gain from an intrinsic data augmentation process and the ability to embed variable-sized windows of audio, and achieve state-of-the-art performance on recent benchmarks \cite{xvec}.

Despite this advantage, we utilize d-vectors for our strategy for their relative simplicity of implementation. d-vectors are still one of the most prominent speaker representation extraction frameworks \cite{park2021review}, so in this initial work we focused on testing  whether we can achieve acceptable  performance with this approach.  \footnote{We are unaware of research that directly compares the performance of x-vectors to d-vectors.}

 When it is possible to cast diarization as a speaker identification task for a predetermined set of speakers, one of the most widely used metrics is Probabilistic Linear Discriminant Analysis (PLDA).  \cite{GarciaRomero2011AnalysisOI}.
 However, in our case, we have unknown speakers, so we use the cosine distance metric to cluster speaker embeddings for simplicity and transparency.

One could also employ a general unsupervised clustering algorithm to perform the speaker verification. Although there have been recent attempts to use deep learning for this purpose, from our initial exploration we found they can be computationally expensive and hard to manipulate. In this work we looked to explore if adequate diarization performance can be achieved in our use-case with a simple non-parametric representation;  more sophisticated models can be a focus of future work.

\section{Speech Embedding Network}

Our method utilizes a speech embedding network trained with the Generalized End-to-End Loss \cite{wan2020generalized} to encode speech into d-vectors \cite{Varianidvec}. This loss function can mimic the process of speaker enrollment and verification during training \cite{heigold2015endtoend}. We build on previous implementations \cite{wang2018speaker}. We use Resemble.AI's pre-trained Resemblyzer, a variant of the \textit{speaker encoder network} \cite{jia2019transfer}, which we refer to as the \textit{Voice Encoder}. These networks follow the general architecture and parameters of the \textit{Speaker Encoder} in \cite{wan2020generalized}. 

Resemble.AI trained their Voice Encoder on VoxCeleb1\&2 \cite{vox1,vox2} as well as the LibriSpeech-other \cite{librispeech} dataset, for a total of $\sim5,600$ speakers in their training data. For comparison, there were $\sim18,000$ speakers used in \cite{wan2020generalized}. Their model is a three-layer LSTM network with a single linear layer and additional rectified linear activation function (ReLU) layer on top of the linear layer, normalizing projections between 0 and 1, as d-vector components fall between -1 and 1. In our fork\footnote{\url{github.com/JeffT13/VoiceEncoder}} of the Resemblyzer we added the capability to retain speaker labels with d-vectors and process audio files that are too large to fit in GPU memory. 

Our audio data is sampled at 16KHz and prepared for the encoder using the processing procedure of the Voice Encoder, which follows \cite{Prabaudioproc}, unless otherwise specified. During Resemble.AI's training of the Voice Encoder, audio files are divided into single speaker utterances. In contrast, when we utilize their model for inference we do not segment the audio file, so our processing unit is the entire audio file. We refer to the entire audio file as the \textit{case} or \textit{proceeding} in this work for clarity. 

First we perform Voice-Activity-Detection (VAD), which yields variable-length segments of voice activity across a case. During inference, we drop non-speech, concatenate the VAD segments into a single sequence and then split the sequence into non-overlapping windows which we refer to as "partial utterances". We convert the entire VAD sequence into a 40-bin mel-spectrogram and process these partial utterances, or range of mel-spectra, through the Voice Encoder. The encoder contains a hyperparameter \code{rate} to determine the size of the partial utterance window, or frequency at which we batch mel-spectra. It is defined as the number of non-overlapping windows per second.

The amount of audio (in time) represented by a single d-vector can vary greatly with the \code{rate}. The maximum width for partial utterance window is 1600ms (set in training), so our \code{rate} is lower bounded to $.625$, producing a single d-vector for the full 1600ms window. The \code{rate} is also upper bounded at $100$, where we are creating one d-vector for every mel-spectrum individually (which we produce in 10ms hops). A larger rate decreases the ratio of d-vectors that would contain speech from two speakers in the label retention process, but increases the computational cost to perform diarization with the d-vectors. We explore the ramifications of the \code{rate} parameter further in Section 5.

\section{Reference Dependent Speaker Verification}

\begin{figure}[ht]
    \captionsetup{justification=centering}
    \centering
    \includegraphics[width=0.33\textwidth]{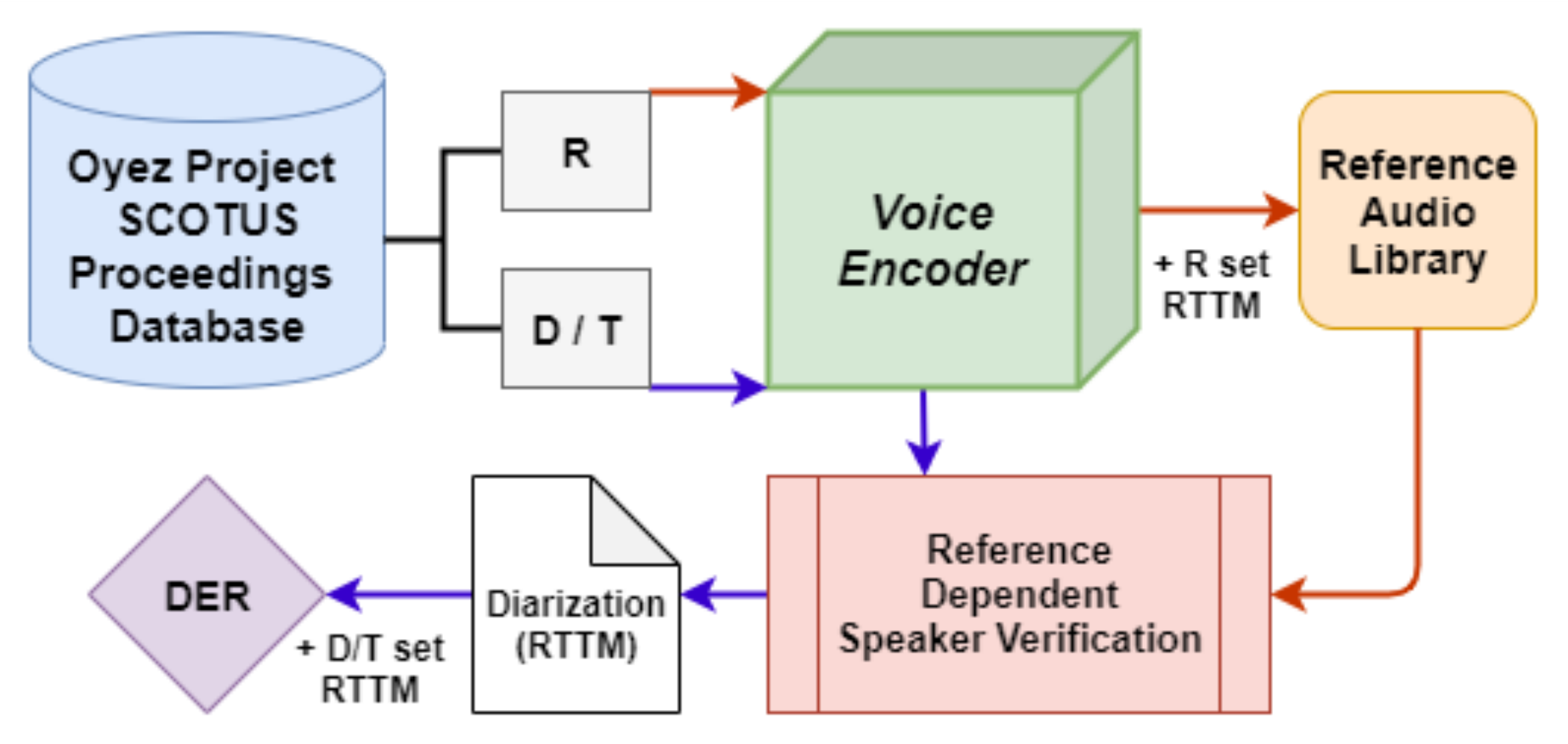}
    \caption{Audio Diarization Pipeline}
    \label{fig:flow}
\end{figure}

The Voice Encoder is the only formal model necessary to perform the Reference Dependent Speaker Verification (RDSV) process for a set of audio recordings, but the procedure also requires some annotated data. Specifically, it requires an annotated Reference set (R), Development set (D) and Test set (T). R is used to generate our Reference Audio Library (RAL), D is used to tune the end-to-end diarization process (i.e. the embedding network and RDSV hyperparameters) and T is used for a final evaluation on heldout annotated data to benchmark the pipeline. A visualization of the pipeline including the data partitions can be seen in Figure~\ref{fig:flow}. Note that the R set (orange arrows) must be processed before diarizing the D or T set (purple arrows). 

The Oyez project \cite{oyez} offers an API to pull the audio for SCOTUS oral arguments and their accompanying transcriptions. This makes generating the dataset splits for our application simple. Note that the SCOTUS oral arguments are a good testbed for our system since almost every judge is present in almost every case. However, a large portion of the audio is the speech of lawyers who will not be present in our RAL.

It is important to pick informative cases for our R set. A case is a good choice for the R set if it contains adequate amounts of speech from target speakers which the RAL does not already contain. Since our focus is on diarizing the judicial speech in the audio, it is crucial to have adequate speech of the judges in the target audio. This is an easy requirement to meet for the SCOTUS, but will be much more challenging for other court systems as they have much larger populations of judges.

\subsection{Reference Audio Library}

The Reference Audio Library (RAL) is the dictionary of speakers found in the R set and the associated d-vectors for those speakers. The SCOTUS oral argument transcriptions are converted to Rich Transcription Time-Marked Files (RTTM) \cite{nist_2009} annotations. Speaking intervals for a given speaker are extracted and a subsequence of d-vectors from the embedded audio file are created if it is above a minimum length parameter \code{min\_audio\_len}. We then take the average of the d-vectors in that subsequence as our representative d-vector for the speaking interval. This process is done for each speaking interval separately, so a speaker will be represented by multiple d-vectors, each d-vector representing a different speaking interval. We also set our RAL to only retain judges as speakers and drop any speaker that does not meet a minimum number of separate speaking intervals to reference, which is a parameter in our implementation \code{min\_ref\_count}. This parameter's only impact is to ensure we meet the minimum duration for each of the Supreme Court judges we plan to diarize.

\begin{figure}[ht]
    \captionsetup{justification=centering}
    \centering
    \includegraphics[height=10cm, keepaspectratio]{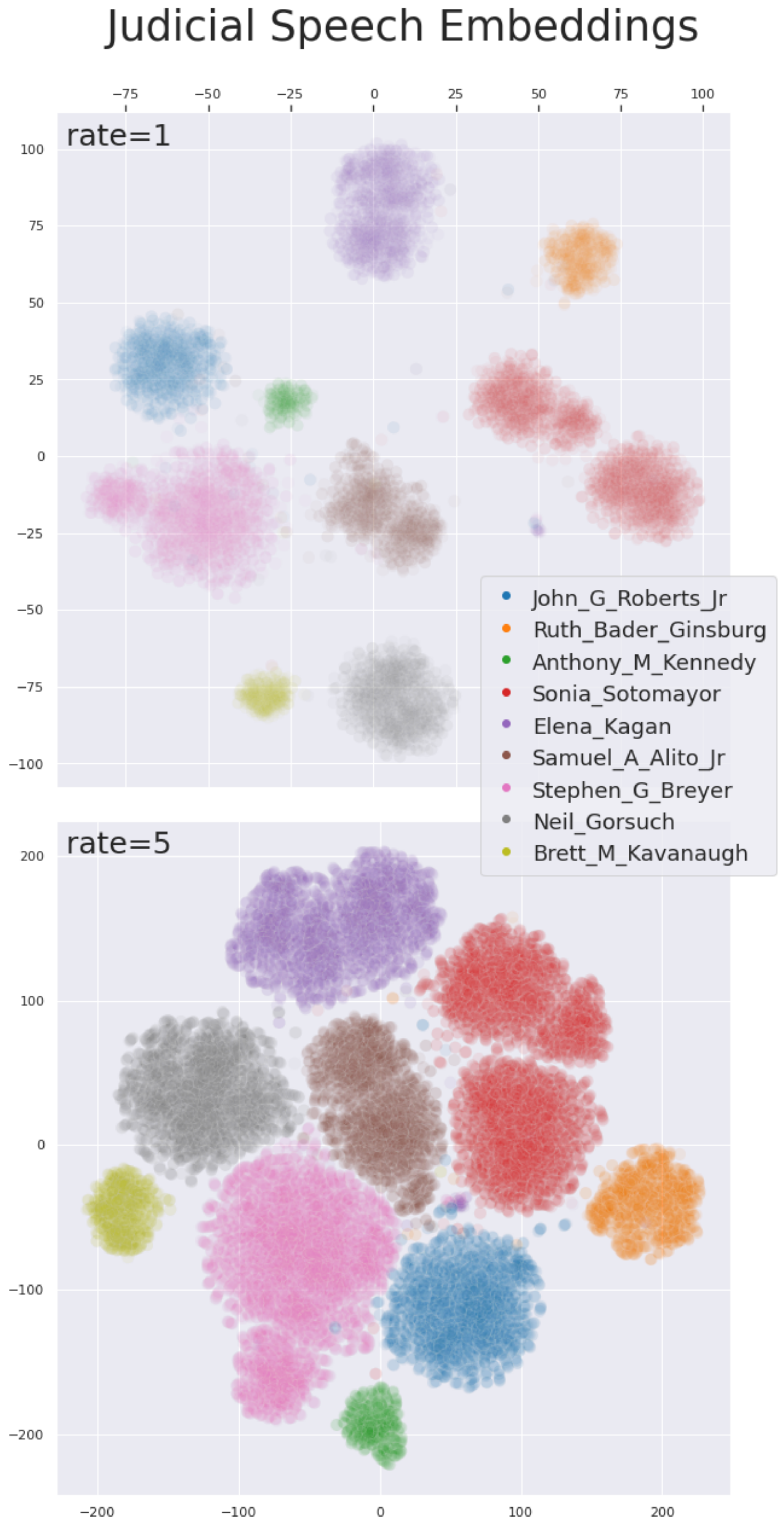}
    \caption{d-vectors of Judicial Speech in R set \\
    for minimum rate (top) and optimal rate (bottom) \\
    Dimensionality Reduction through t-SNE}
    \label{fig:VEboth}
\end{figure}

\subsection{Speaker Label Inference}

Our objective in RDSV is to use the RAL embeddings for each judge to infer who is speaking at each time step in an unseen case. We do this using the cosine distance metric \cite{ivec}. Our implementation creates an affinity matrix between the reference embeddings and the case, storing the dot product between each reference embedding and each time step of the continuous case embedding. Because d-vectors have unit norm, this amounts to the cosine similarity between the reference audio and target audio per judge. The speaker with the reference embedding that scores the highest similarity to the embedded target audio at a particular time step is chosen as the inferred speaker.

If this were a closed-set Speaker Identification task, this would be sufficient. But since we assume our RAL will not account for every speaker, the diarization process includes two threshold hyperparameters, \code{sim\_thresh} and \code{score\_thresh}, used to label unrecognized voices. If the highest similarity score between a speaker reference and the target audio segment is under \code{score\_thresh} and the difference between that similarity score and the highest similarity score between the same target audio segment and the reference of any other speaker (judge) is above \code{sim\_thresh}, then we label this an unreferenced speaker. 

\begin{figure*}[h!]
    \captionsetup{justification=centering}
    \centering
    \includegraphics[width=.73\textwidth, keepaspectratio]{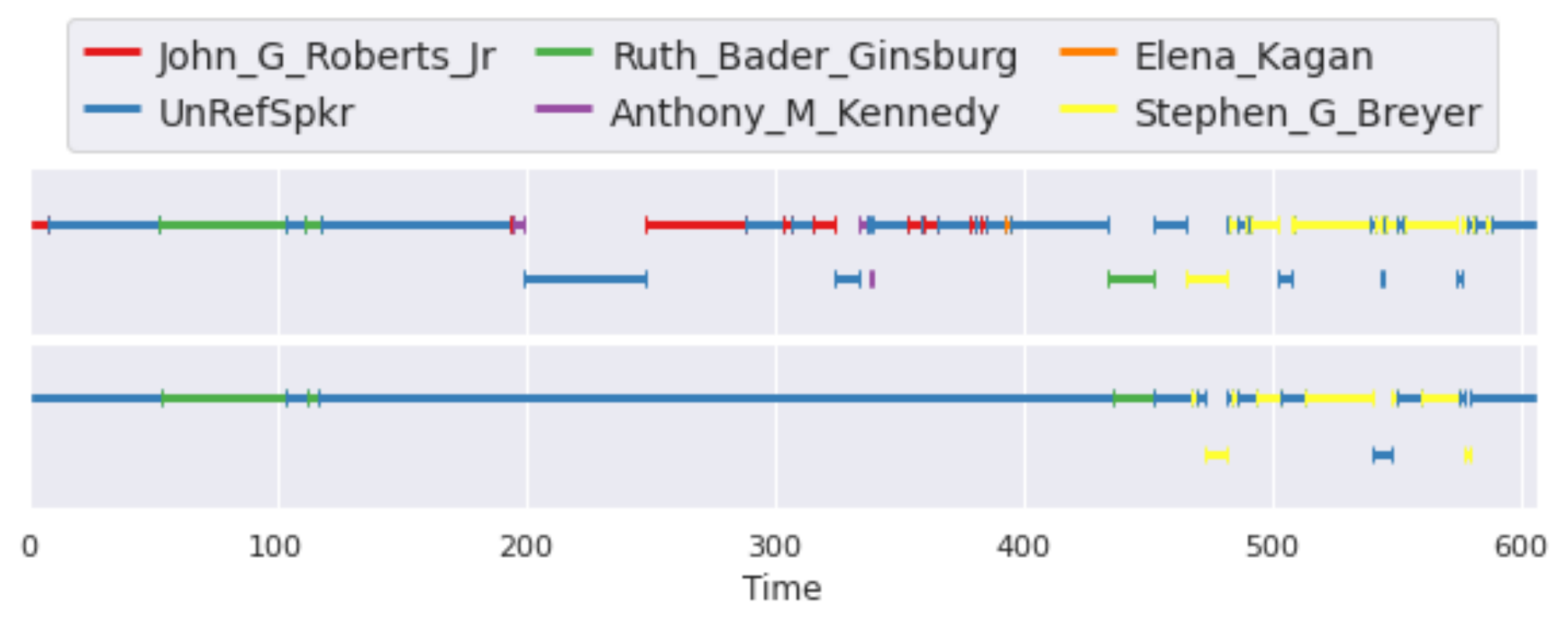}
    \caption{Groundtruth (top) and Inferred (bottom) Segmentation Plots for the first 10 minutes of a case from the test set.}
    \label{fig:seg}
\end{figure*}

\section{Experiments}

\subsection{Supreme Court Audio} 

We evaluate our pipeline on the SCOTUS oral arguments and transcriptions database, accessed through the Oyez project \cite{oyez}. We pull all cases (with transcriptions) heard in one session from Supreme Court docket terms 16-18\footnote{docket terms represent calendar years for the SCOTUS and group hearings by their date (i.e. docket 16 is Oct 2016-Sept 2017)} . We allocate 3 unique cases from each docket 16 and 18 (6 total) into the both R and D set (12 total), which are used to analyze the Speech Embedding network and tune our diarization pipeline. We then allocate 25 cases from docket 17 as our test set, which are 58 minutes on average and are about 51 minutes each after VAD.

\subsection{Speech Embedding Network Analysis}

Because both speaker label quality and computational cost are impacted by the \code{rate} parameter, we test the quality of the Voice Encoder speaker representations for \code{rate}=$[1,3,5,7]$. To do this, we encode our R and D set for each rate and perform a multiclass one-vs-rest Logistic Regression with scikit-learn \cite{scikit-learn} to infer d-vector speaker labels. We train the model on only the d-vectors from the judicial speech found in the R set audio. We then evaluate the ensemble classifier's performance on the d-vectors from the full proceedings in our D set. We do this by thresholding the predicted probability from the classifier to predict a non-judge speaker at $[.85, .9, .95]$. We set our \code{rate}=$5$ for our diarization process as we found that a rate of $5$ had the most stable accuracy as the threshold increased and the highest overall accuracy.

We also visualize the d-vectors representing judicial speech from the cases in our R set for our lowest and optimal rate. We reduce our d-vectors to two dimensions using sklearn's t-distributed Stochastic Neighbor Embedding (tSNE) with a perplexity of 50 (default parameters otherwise). We then plot the shrunken d-vectors with their speaker labels to see how well our embeddings tend to cluster and if our d-vector labels are accurate. With the help of Figure~\ref{fig:VEboth} we can see that even our lowest rate performs reasonably, but there is significantly higher ratio of dispersed d-vectors in comparison to the higher rate. This supports our conclusion to use a higher rate rather than the minimum for our diarization pipeline.

\subsection{Supreme Court Diarization}

After embedding the R and D sets we tune our diarization process hyperparameters. We validate across the ranges \code{score\_thresh}=$[.75, .8, .85, .9]$ and \code{sim\_thresh}=$[.05, .075, .1, .125]$. To evaluate performance during tuning we use the DiarizationErrorRate (DER) in pyannote.metrics \cite{pyann} with a collar of half a second, as most of the speaker turns tend to be on the longer side.  We track the Average DER ($\mu_{D}$) and DER standard deviation ($\sigma_{D}$) across all cases in the T set. Because our diarization process assigns all unreferenced speakers to a constant label, we also convert the groundtruth RTTM labels such that all speakers not members of the RAL get cast to a single label. Our primary metric for performance is $\mu_{D}$, and we use the parameters that minimize the $\mu_{D}$ of the D set in processing our heldout test set. The optimal values found were \code{score\_thresh}$=.85$ and \code{sim\_thresh}$=.1$. Table~\ref{tr} shows the diarization performance of our pipeline on the 25 cases in the test set, including the average amount of audio per case (AC) and average amount of audio after VAD (ACV) in minutes.

\begin{table}[ht]
    \captionsetup{justification=centering}
    \caption{Test Set Diarization Statistics (n=25) \\
        $\mu_{D}$ = Average DER | $\sigma_{D}$ = DER Standard Deviation \\
        AC=Audio per case | ACV=Audio (post-VAD)  per case}
    \centering
    \begin{tabular}{llllll}
        \toprule
        \cmidrule(r){1-2}
        Metric  & $\mu_{D}$ & $\sigma_{D}$ & max DER &  AC & ACV \\
        \midrule
        Value &  13.8 \% & 3.5\% & 19.8\% & 58.9 & 51.36\\
        \bottomrule
    \end{tabular}
    
    \label{tr}
\end{table}

Figure~\ref{fig:seg} shows the output of our system for the first 10 minutes of a typical diarized test case. It includes our prediction (bottom) and the ground truth (top) with unreferenced speakers in our RAL converted to a single identifier. This short segment highlights the large ratio of lawyer speech to judicial speech. We reiterate that our focus is identifying the segments of judicial speech, so the diarization of lawyer speech is out of scope for this paper. We can see that our method does diarize our pre-referenced judges, and does quite well when the speaking segment is not too short, but occasionally struggles to identify a speaker, especially on short segments. This is particularly the case if a judge sounds similar to a lawyer or if limited references were available for that judge compared to the others (which is the case for Justice John G. Roberts).

\section{Discussion \& Summary}

Our approach is an attempt to illustrate the usability of an audio processing pipeline that is domain-specific, exploiting simple, open-source, transparent implementations. By leveraging a pre-trained open-source model in Resemble.AI's \textit{Voice Encoder} and computing cosine similarities for speaker verification, we achieve strong diarization performance on a real-world dataset with relatively uncommon characteristics: an average Diarization Error Rate of $13.8\%$, and a low variance and maximum diarization error rate of $19.8\%$\footnote{We intend to improve diarization performance on short speaking segments before applying our pipeline widely.}.

Our model is parsimonious; both the encoder and speaker verification scoring could evolve in complexity. Both x-vectors as speaker representations and using a neural network to learn the distance metric are the current state-of-the-art, and we look to apply these in future work. We also can consider commercial APIs (Google, IBM, etc.) for speech segmentation, as they seem to perform well.  We believe better speech segment boundaries would help our RDSV pipeline correctly identify short speaking segments from speakers who sound similar to a more common speaker in the audio, which is a challenging situation for our pipeline.

We hope we have encouraged further research into the diarization of legal proceedings. Oral argument transcript text may shed light on a broad spectrum of judicial behaviors and biases, including race-based discrimination \cite{steffensmeier2001judges}, court prediction \cite{kaufman2019improving}, and partisan political bias \cite{harris2019bias}. Professionally, the legal industry relies on enormous corpora of indexed text, primarily of court decisions, that are largely hidden behind paywalls despite being public record. By making diarized transcripts publicly available, we can increase access to public records for those typically locked out of the legal discipline.

\bibliographystyle{IEEEtran}
\bibliography{mybib}

\end{document}